\documentclass[12pt]{article}

\usepackage{epsfig}
\usepackage{amssymb}
\usepackage{subfigure}
\usepackage{graphicx}
\usepackage{color}
\usepackage{jheppub} 

\begin{document}

\baselineskip 6mm
\renewcommand{\thefootnote}{\fnsymbol{footnote}}


\newcommand{\nc}{\newcommand}
\newcommand{\rnc}{\renewcommand}



\newcommand{\tcb}{\textcolor{blue}}
\newcommand{\tcr}{\textcolor{red}}
\newcommand{\tcg}{\textcolor{green}}


\def\ba{\begin{array}}
\def\ea{\end{array}}
\def\be{\begin{eqnarray}}
\def\ee{\end{eqnarray}}
\def\nn{\nonumber\\}


\def\ct{\cite}
\def\la{\label}
\def\eq#1{(\ref{#1})}


\def\a{\alpha}
\def\b{\beta}
\def\g{\gamma}
\def\G{\Gamma}
\def\d{\delta}
\def\D{\Delta}
\def\e{\epsilon}
\def\et{\eta}
\def\ph{\phi}
\def\Ph{\Phi}
\def\ps{\psi}
\def\Ps{\Psi}
\def\k{\kappa}
\def\l{\lambda}
\def\L{\Lambda}
\def\m{\mu}
\def\n{\nu}
\def\th{\theta}
\def\Th{\Theta}
\def\r{\rho}
\def\s{\sigma}
\def\S{\Sigma}
\def\ta{\tau}
\def\o{\omega}
\def\O{\Omega}
\def\pr{\prime}


\def\half{\frac{1}{2}}
\def\goto{\rightarrow}

\def\na{\nabla}
\def\grad{\nabla}
\def\curl{\nabla\times}
\def\div{\nabla\cdot}
\def\pa{\partial}
\def\fr{\frac}

\def\bra{\left\langle}
\def\ket{\right\rangle}
\def\lb{\left[}
\def\lc{\left\{}
\def\ls{\left(}
\def\lp{\left.}
\def\rp{\right.}
\def\rb{\right]}
\def\rc{\right\}}
\def\rs{\right)}

\def\vac#1{\mid #1 \rangle}


\def\td#1{\tilde{#1}}
\def\check{ \maltese {\bf Check!}}


\def\Tr{{\rm Tr}\,}
\def\det{{\rm det}}
\def\text#1{{\rm #1}}


\def\bc#1{\nnindent {\bf $\bullet$ #1} \\ }
\def\ch {$<Check!>$ }
\def\ss {\vspace{1.5cm}}
\def\inf{\infty}

\begin{titlepage}

\hfill\parbox{2cm} { }

 
\vspace{2cm}

\begin{center}
{\Large \bf Time-dependent quantum correlations \\
in two-dimensional expanding spacetime }

\vskip 1. cm
   {Chanyong Park$^{a}$\footnote{e-mail : cyong21@gist.ac.kr}}

\vskip 0.5cm

{\it $^a$ Department of Physics and Photon Science, Gwangju Institute of Science and Technology,  Gwangju  61005, Korea}

\end{center}

\thispagestyle{empty}

\vskip2cm


\centerline{\bf ABSTRACT} \vskip 4mm

\vspace{1cm}

In expanding universes, the entanglement entropy must be time-dependent because the background geometry changes with time. For understanding time evolution of quantum correlations, we take into account two distinct holographic models, the dS boundary model and the braneworld model. In this work, we focus on two-dimensional expanding universes for analytic calculation and comparison. Although two holographic models realize expanding universes in totally different ways, we show that they result in the qualitatively same time-dependence for eternal inflation. We further investigate the time-dependent correlations in the radiation-dominated era of the braneworld model. Intriguingly, the holographic result reveals that a thermal system in the expanding universe is {\it dethermalized} after a critical time characterized by the subsystem size.

\vspace{2cm}

\end{titlepage}

\renewcommand{\thefootnote}{\arabic{footnote}}
\setcounter{footnote}{0}

\tableofcontents


\section{Introduction}

For  understanding strongly interacting systems at a low energy scale, it is important to figure out the energy scale dependence of physical systems via a nonperturbative renormalization group (RG) flow. At present, unfortunately, there is no well-established nonperturbative and analytic tool. In this situation, there was an interesting proposal in the string theory, the so-called anti-de Sitter (AdS)/ conformal field theory (CFT) correspondence or holography \cite{Maldacena:1997re,Gubser:1998bc,Witten:1998qj,Witten:1998zw}. The holography claims that a strongly interacting conformal field theory (CFT) maps to a one-dimensional higher dual gravity theory. Though it is a formidable task to account for a nonperturbative quantum field theory (QFT), the AdS/CFT proposal provides a new chance to uncover nonperturbative aspects of strongly interacting systems. Due to this reason, people widely exploited the holographic technique in studying nonperturbative features of phase transition, linear response, and quantum entanglement. In this work, we investigate time-dependent quantum correlations, the entanglement entropy and a two-point correlation function, in expanding universes \cite{Koh:2020rti,Park:2020jio}

One of the important quantities revealing quantum features is the entanglement entropy \cite{Calabrese:2004eu,Calabrese:2009qy,Lewkowycz:2013nqa}. The entanglement entropy represents the quantum correlation between two distinct subsystems. In general, it is not easy to calculate the entanglement entropy of an interacting QFT. Based on the holography, Ryu and Takayanagi (RT) proposed that the entanglement entropy of QFT is associated with the area of the minimal surface extending to the dual geometry \cite{Ryu:2006bv,Ryu:2006ef}. Recent studies on the entanglement entropy showed that the RT proposal is very successful. For example, a two-dimensional CFT is special because it has infinitely many symmetries and the modular invariance. These large symmetries enable us to calculate the entanglement entropy exactly \cite{Calabrese:2004eu,Calabrese:2009qy}. When applying the RT formula, intriguingly, it was shown that the holographic calculation exactly reproduces the results of a two-dimensional CFT \cite{Ryu:2006bv,Ryu:2006ef,Nishioka:2009un}. The RT proposal was further applied to higher-dimensional cases and showed that the entanglement entropy can characterize important quantum properties like $a$- and $F$-theorems \cite{Zamolodchikov:1986gt,Myers:2010xs,Myers:2010tj}.

When applying the RT formula, we have to consider a space-like hypersurface at a given time and divide the boundary space into two subsystems. If the dual geometry is static, taking a space-like hypersurface looks natural because of the time translation symmetry. In this case, the resulting entanglement entropy becomes time-independent. If we want to know the time-dependence of the entanglement entropy, we must break the time translation symmetry. One way to break the time translation symmetry is to consider a time-dependent geometry. In this case, we cannot directly apply the RT formula because of the absence of the time translation symmetry. It was argued that one must exploit a covariant or Hubeny-Rangamanni-Takayanagi (HRT) formula, instead of the RT formula, in the time-dependent background geometry \cite{Hubeny:2007xt}. Although the concept of the HRT formula is concrete, it is not easy to evaluate such a time-dependent entanglement entropy exactly even in the holographic setup. In this work, we take into account a three-dimensional AdS space as a toy model, whose boundary is given by a dS space representing two-dimensional eternal inflation. On this time-dependent background geometry, we calculate the time-dependent entanglement entropy exactly by applying the HRT formula beyond the RT formula.  

Except for the two-dimensional inflationary universe, the dS boundary model cannot realize other universes expanding by power-laws. In order to study the power-law expansion, therefore, we take into account another holographic model called the braneworld model \cite{Randall:1999ee,Randall:1999vf,Chamblin:1999ya,Chamblin:1999by,Park:2000ga}, where we consider a moving brane in an asymptotic AdS geometry. The braneworld model allows us to describe the universes expanding by power-law holographically \cite{Park:2020jio}. Since the background geometry of the braneworld model is time-independent, the RT formula in the braneworld model describes time-dependent quantum correlations. From now on, we concentrate on a two-dimensional braneworld model which can be easily generalized to the higher-dimensional case. The two holographic models, the dS boundary and braneworld models, describe the expanding universes in totally different ways. Despite this fact, we show that these two holographic models yield a similar time-dependent entanglement entropy for an inflationary universe. We also investigate a radiation-dominated universe in the braneworld model. In this case, intriguingly, the time-dependent entanglement entropy reveals a remarkable feature. A black hole geometry, in general, allows the thermodynamic interpretation. The Bekenstein-Hawking entropy maps to the thermal entropy of the dual QFT. In the IR regime, it has been shown that the entanglement entropy approaches the thermal entropy \cite{Liu:2012eea,Hartman:2013qma,Liu:2013iza,Kim:2016jwu}. In the radiation-dominated universe, we show the expansion of the universe prevents radiations from being thermalized which we call '{\it dethermalization}'.

The entanglement entropy represents the quantum correlation between two macroscopic subregions,  while a two-point function of a local operator describes a microscopic quantum correlation \cite{Susskind:1998dq,Balasubramanian:1999zv,Park:2020nvo}. From the holographic viewpoint, these two distinct quantum correlations are realized by geometrical objects. For a three-dimensional dual gravity theory, in particular, they are described by the same geodesic curve. This fact may indicate a  relation between the macroscopic and microscopic correlations. To see the connection between them, we investigate the time-dependent two-point function of local operators in expanding universes. In the expanding universe the entanglement entropy increases with time, while the microscopic two-point function shows rapid suppression. This is because the increase of the entanglement entropy in the medium may yield a strong screening effect on the microscopic two-point function.

The rest part of this paper organizes as follows. In Sec. 2, we first review the dS boundary model and then investigate the macroscopic and microscopic correlations in an eternally inflating universe. In Sec. 3, We discuss an inflationary universe appearing in the braneworld model and look into the time dependence of the macroscopic and microscopic correlations. In Sec. 4, we further investigate the quantum correlation in the radiation dominated era of the braneworld model. We close this work with some concluding remarks in Sec. 5.


\section{Covariant entanglement entropy in dS space}

We take into account a two-dimensional inflationary universe. In the holographic setup, there are two distinct models realizing an inflationary universe. The first one considers a three-dimensional AdS space whose boundary is given by a dS space. For later convenience, we call this model a dS boundary model \cite{Maldacena:2012xp,Fischler:2013fba,Chu:2016pea,Koh:2020rti}. The second describes a moving brane in the AdS space. This model was known as the braneworld or Randall-Sundrum model \cite{Randall:1999ee,Randall:1999vf,Park:2000ga,Park:2020jio}. In this section, we first discuss the dS boundary model and investigate the time-dependent correlations.

Relying on the boundary topology, there are several representations of a three-dimensional AdS space. In the global patch, the AdS space with a dS boundary is expressed as
\be		 
ds^2 = \fr{R^2 d z^2}{z^2 (1+z^2/R^2)}  + \fr{R^2}{z^2} \lb - d t^2 + R^2 \cosh^2 (t/R)  \ 
d \th^2 \rb ,
\ee
where $d \th^2 $ means a metric on a unit circle. To characterize the boundary spacetime manifestly, we assume that the boundary is located at a fixed radial position $z=\e$. Introducing a cosmological time $\ta$ defined as $\ta=R t /\e$, the boundary metric reduces to the Friedmann-Lema\^{i}tre-Robertson-Walker (FLRW) type metric \cite{Kinney:2009vz,Baumann2018}
\be		 
ds^2 =  - d \ta^2 + \fr{ \cosh^2 (H \ta) }{H^2} \ d \th^2 ,
\ee
which is the metric of a two-dimensional dS space with a Hubble constant $H = \e/R^2$. According to the AdS/CFT correspondence, this AdS geometry is dual to a CFT living in a two-dimensional dS space.

In the Poincare patch, on the other hand, the AdS space with a dS boundary can be represented as
\be		 	\la{metric:PoincareAdS}
ds^2 = \fr{R^2 d z^2}{z^2 (1+z^2/R^2)}  + \fr{R^2}{z^2} \lb - d t^2 + e^ {2 t/R}  \ 
d x^2  \rb ,
\ee
where $d x^2$ indicates the metric of a one-dimensional straight line. At the boundary with $z=\e$, the AdS metric reduces to 
\be		 	
ds_B^2 =  \fr{R^2}{\e^2} \lb - d t^2 + e^ {2 t/R}  \  d x^2  \rb ,
\ee
which is nothing but the metric of a two-dimensional dS space with a flat spatial section. In terms of the cosmological time, the boundary metric reduces to 
\be		 \la{metric:inducedmetric}
ds_B^2 =    - d \ta^2 + \fr{e^ {2 H \ta} }{H^2 R^2}   \  d x^2   ,
\ee
where the Hubble constant is again given by $H=\e/R^2$. If we introduce a new time coordinate $\ta'$ with $\ta' = \ta - \log \ls HR \rs /H$,  the boundary metric further reduces to a simple FLRW metric
\be			\la{result:FLRWmetric}
ds_B^2 =  - d \ta'^2 + e^ {2 H \ta'}  \  d x^2  ,
\ee
This is the well-known dS metric representing eternal inflation. Now, we assume that the dS space starts at $\ta=0$, for simplicity, and that the universe, before $\ta=0$, is in the Hartle-Hawking vacuum state satisfying the no boundary condition \cite{Hartle:1983ai}. Then, the eternal inflation allows the time range of $0\le \ta < \infty$. It is worth noting that the bulk metric is invariant under the constant shift of time and rescaling of $x$
\be
\ta \to \ta' = \ta + a \quad {\rm and} \quad x \to x' = e^{-H a} x  .
\ee 
Due to this isometry, the shift of $\ta$ does not give any significant effect. Moreover, the metric \eq{metric:inducedmetric}, without loss of generality, reduces to \eq{result:FLRWmetric} by redefining the spatial coordinate $x$.

\subsection{Entanglement entropy in the dS boundary model}

Now, we consider the time-dependent entanglement entropy in the inflating two-dimensional universe. In general, it is not easy to calculate the entanglement entropy of interacting quantum field theories. Even in this case, the holographic technique based on the AdS/CFT correspondence may be helpful. To calculate the entanglement entropy holographically, the authors of Ref. \cite{Ryu:2006bv,Ryu:2006ef} proposed that the entanglement entropy is associated with the area of the minimal surface extending to the dual geometry. This is the story when the bulk metric is time-independent. If one considers a time-dependent spacetime like an expanding universe, one must exploit another method called HRT formula, instead of the RT formula \cite{Hubeny:2007xt}. Although the HRT formula was well established conceptually, it is still difficult to calculate the time-dependent entanglement entropy  exactly even in the holographic setup. In this situation, it was argued that the RT formula in a small subsystem size limit leads to the leading contribution to the HRT formula \cite{Koh:2020rti}. In this section, we investigate the entanglement entropy of the HRT formula, beyond the RT formula. Note that the exact calculation of the HRT formula is possible only for a two-dimensional inflationary universe because the HRT formula is solvable only in this case. 

For later convenience, we introduce a new time coordinate 
\be
T = R \, e^{-t/R} ,
\ee
which corresponds to the conformal time of the boundary dS space. Then, the bulk AdS metric is rewritten as
\be		 
ds^2 = \fr{R^2}{z^2} \lb \fr{d z^2}{1+z^2/R^2}  + \fr{R^2}{T^2} \ls - d T^2 + d x^2  \rs  \rb .
\ee
In this case, the range of $T$ is restricted to $0 \le T \le T_i =R$ where $T_i$ corresponds to the initial time. On this  background geometry, let us take into account the holographic entanglement entropy described by the HRT formula. On the dual field theory side, this corresponds to a time-dependent entanglement entropy because the boundary space changes in time. To calculate the entanglement entropy, we divide the boundary space into two parts, a subsystem and its complement, and parameterize the subsystem as $-l/2 \le x \le l/2$. Regarding $z$ and $T$ as functions of $x$, the HRT formula gives rise to the following entanglement entropy
 \be			\la{action:HEEAdS3}
S_E = \fr{1}{4 G} \int_{-l/2}^{l/2} dx \ \fr{R}{z} \sqrt{\fr{z'^2}{1 + z^2 /R^2} + \fr{R^2}{T^2} \ls 1 - T'^2  \rs} \ ,
\ee
where the prime indicates a derivative with respect to $x$. This action governs a minimal surface extending to the time-dependent bulk geometry. It was shown that the equations of motion  of a minimal surface is solvable for a three-dimensional static AdS space with a planar boundary \cite{Chu:2016pea}. In the present work we take into account an expanding universe, so that we should be careful to impose the boundary conditions.

Before finding an analytic solution, let us first think of natural boundary conditions that the minimal surface must satisfy. To calculate the entanglement entropy we divided the boundary space into two parts, so that the border of two subsystems called the entangling surface is located at the boundary ($z=0$). Since the minimal surface we want to find must anchor to the entangling surface, the solution has to satisfy $z(\pm l/2) = 0$. This is one of the natural boundary conditions. Noting that the above entanglement entropy is invariant under $x \to - x$, this fact together with smoothness of the minimal surface requires that $z'$ and $T'$ have to vanish at $x=0$. Denoting the values of $z$ and $T$ at $x=0$ as $z_0$ and  $T_0$, these two values characterize the turning point of the minimal surface. In this case, the minimal surface extends only in the range of $0 \le z \le z_0$ and $T_0 \le T \le T_B$ where $T_B$ is the time measuring the entanglement entropy at the boundary. Therefore, the configuration of the minimal surface should also satisfy three more  boundary conditions, $z'=T'=0$ at $x=0$ and $T=T_B$ at $x=\pm l/2$. These boundary conditions uniquely determines the minimal surface's configuration, as will be shown later. 

To find the minimal surface satisfying all boundary conditions discussed before, we first look into the equations governing the shape of the minimal surface. Since the above entanglement entropy depends on $x$ implicitly, there exists one conservation law which appears as a constraint 
\be
\frac{\sqrt{R^2+z^2}}{ \sqrt{\left(1-T'^2\right) \left(R^2+z^2\right)+T^2  z'^2}} 
= \frac{T z}{ T_0 \ z_0}  .
\ee
We can also find two more dynamical equations which are not independent. Combining these equations, we finally obtain a simple dynamical equation of $T$ \cite{Chu:2016pea}
\be
0 = T T'' + T'^2 - 1 .
\ee
A general solution of this dynamical equation is given by
\be
T = \sqrt{  (x + c_1)^2  - \fr{c_2}{4}} ,
\ee
where $c_1$ and $c_2$ are two integral constants. Imposing one of the natural boundary conditions, $T'=0$ at $x=0$, fixes the value of one integral constant to be $c_1 =0$. The other natural boundary condition, $T=T_B$ at $x=\pm l/2$, further fixes the remaining integral constant to be 
\be  	\la{result:constant}
c_2  =  l^2 - 4 T_B^2   .
\ee 
Using these results, the value of $T_0$ at the turning point is given by $T_0 =\sqrt{4 T_B^2 - l^2}/2$ and the time on the minimal surface is finally determined as a function of $x$ 
\be			\la{Solution:tdirection}
T (x) = \sqrt{  T_0^2 + x ^2  } .
\ee
The time difference between $T_0$ at the turning point and $T_B$ at the boundary is given by
\be
\D T \equiv T_B - T_0 = T_B  \ls 1 - \sqrt{1 - \fr{l^2}{4 T_B^2}}  \rs .
\ee
We can easily check that $T'$ has a finite value at the boundary
\be
T' \ls \pm \fr{l}{2} \rs = \fr{l}{2T_B} .
\ee
In the small subsystem size limit with $l / T_B \ll1$, since $T'$ and $\D T$ vanish, the effect of $T'$ in \eq{action:HEEAdS3} is negligible. In other words, the RT formula for $l/T_B \ll 1$ gives rise to the leading contribution to the HRT formula as mentioned in Ref. \cite{Koh:2020rti}.

Using the above solution $T$, $z'$ reduces to a function of $ z$ and $x$
\be
z ' =  \frac{T_0\sqrt{R^2+z^2} \sqrt{  z_0^2 - z^2  }}{z \left(T_0^2 + x^2 \right)}  ,
\ee
and its solution is given by
\be
z (x) = \sqrt{z_0^2 \cos^2 \left(c_3+\tan^{-1}\left(\frac{x}{T_0}\right)\right) -R^2 \sin^2 \left(c_3+\tan ^{-1}\left(\frac{x}{T_0}\right)\right)}  ,
\ee
with an additional integral constant $c_3$. Imposing the  boundary condition, $z(\pm l/2) = 0$, the integral constant is fixed to be
\be
c_3 = \tan ^{-1}\left(\frac{z_0}{R}\right)-\tan ^{-1}\left(\frac{l}{2 T_0}\right) .
\ee
Since all integral constants have already been determined, the last remaining boundary condition, $z'=0$ at $x=0$, determines $z_0$ as a function of the other parameters 
\be         \la{Result:tpvssize}
z_0 = \fr{l R}{\sqrt{4 T_B^2 - l^2 } } .
\ee
Finally, the solution $z$ satisfying all boundary conditions reduces to the following simple form
\be		\la{Solution:zdirection}
z(x) = \fr{R \sqrt{l^2 - 4 x^2}}{\sqrt{ 4 T_B^2 - (l^2 - 4  x^2)}}  .
\ee
As a consequence, the configuration of the minimal surface is analytically determined by \eq{Solution:tdirection} and \eq{Solution:zdirection}, which are the exact solution of the HRT formula. In the small subsystem size limit, the range of $x$ satisfies the following inequality, $l^2 - 4 x^2 \le l^2 \ll T_B^2$, so that the leading behavior of $z$ is well approximated by
\be			\la{result:solofRTformula}
z(x) = \fr{R \ \sqrt{l^2 - 4 x^2}}{  2 T_B } ,
\ee
which is nothing but the solution of the RT formula at a given time $T_B$.

Now, we evaluate the time-dependent entanglement entropy by using the obtained solutions. Substituting the solutions into the HRT formula, we obtain the following integral
\be
S_E = \fr{R l}{G}   \int_0^{l/2 - \d}  \fr{dx}{   l^2 - 4 x^2  }  = \fr{R}{4 G} \log \fr{l}{\d} - \fr{R \d}{4 G l} + {\cal O} \ls \d^2 \rs ,
\ee
where $\d$ corresponds to an appropriate UV cutoff in the $x$-coordinate. Rewriting $\d$ in terms of the UV cutoff $\e$ in the $z$-coordinate
\be
\d = \fr{T_B^2  \e^2}{l R^2}  \ls 1 + \fr{T_B^2 - l^2}{l^2 R^2} \e^2 + {\cal O} \ls \e^4\rs \rs ,
\ee
the resulting entanglement entropy in terms of $\e$ becomes
\be
S_E = \fr{c_{CFT}}{3}  \log \ls \fr{l \ R}{ \e \ T_B} \rs  + \fr{c_{CFT}}{6}  \ls 1 - \fr{2 T_B^2}{l^2 } \rs  \fr{\e^2}{R^2}  + {\cal O} \ls \e^2 \rs  ,
\ee
where the central charge of the dual CFT is given by $c_{CFT} = \fr{3R}{2 G}$ \cite{Balasubramanian1999}. Note that $T_B$ is the conformal time measured at the boundary. To understand the time dependence of the entanglement entropy in the expanding universe, we need to represent the entanglement entropy in terms of the cosmological time which is related to the conformal time by
\be
T_B = \fr{e^{-H \ta}}{H}  .
\ee
Then, the boundary metric of the AdS space reduces to \eq{result:FLRWmetric} and the entanglement entropy is rewritten as
\be
S_E = \fr{c_{CFT}}{3}  \log \ls \fr{l \, e^{H \ta}}{ R } \rs  + \fr{c_{CFT} \, H^2 R^2}{18} \ls 1 - \fr{2    e^{-2 H \ta} }{ H^2 l^2 } \rs  + {\cal O} \ls  e^{-4 H \ta} \rs .
\ee

In a time-dependent geometry, we can define two distinct distances. From the boundary theory viewpoint, the subsystem size $l$ corresponds to a comoving distance which is measured in the comoving frame. The comoving distance is not time-dependent. We also define a physical distance by multiplying the scale factor
\be
D =  l e^{H \ta}  .
\ee
The physical distance usually changes with time because the background geometry is time-dependent. The above holographic result shows that the entanglement entropy in the late time era ($H \ta \gg 1$) increases linearly with time  
\be			\la{result:leadngHEE}
S_E \approx \frac{c_{CFT} }{3}  H   \tau .
\ee
This is because the physical subsystem size increases exponentially during the eternal inflation. 

Now, let us consider how the $c$-function behaves in the inflationary universe. Defining a $c$-function as $c \equiv 3  \, \pa S_E / \pa \log D$, the time evolution of the $c$-function in the late time era ($D/l \gg 1$) is given by
\be		\la{result:cfunctionindSmodel}
c(\ta) = 3 \fr{\pa S_E}{\pa \log D} = c_{CFT}  + \fr{2 c_{CFT}}{3}   \fr{  R^2}{l^2}  e^{- 2 H  \ta}    + {\cal O} \ls e^{-4 H  \ta} \rs  .
\ee
This result shows that, as time elapses, the $c$-function exponentially suppresses and finally saturates $c=c_{CFT}/3$. The monotonic decrease of the $c$-function looks similar to the well-known $c$-theorem that the $c$-function monotonically decreases along the RG flow. If there exists a relation between $c$-theorems evolving in time and along the RG flow, then we may identify the direction of time with the direction of the RG flow \cite{Boer2001,Skenderis2002}. This relation may provide a clue for understanding why irreversibility appears in the macroscopic system even after the unitary time evolution of reversible microscopic quantum systems \cite{Park:2015hcz,Kim:2016jwu}.

\subsection{Two-point function of a local operator during inflation}

In the previous section, we studied the holographic time-dependent entanglement entropy during the eternal inflation. The entanglement entropy gives us information about the quantum correlation between two macroscopic subregions. From the more fundamental viewpoint, this macroscopic quantum correlation may be associated with the sum of all microscopic quantum correlations. In this section, we discuss the connection between the macroscopic entanglement entropy and microscopic two-point functions in the eternal inflationary cosmology.

In the holographic study, it was argued that the geodesic curve connecting two local operators  represents an equal time two-point function \cite{Susskind:1998dq}
\be    		\la{Formula:two-pt function}
\bra O (x) \, O (x') \ket \sim e^{-m L(|x-x'|)}  ,
\ee
where $L(|x-x'|))$ denotes a geodesic length extending to the dual geometry. This holographic two-point function is valid in a $(d+1)$-dimensional asymptotic AdS geometry, regardless of the dimension $d$. In the holographic study, the entanglement entropy of a $d$-dimensional CFT is governed by a $(d-1)$-dimensional minimal surface extending to a $(d+1)$-dimensional AdS space. On the other hand, the geodesic length governing the two-point function is described by a one-dimensional curve. Due to the different dimensions, the geodesic length is usually different from the minimal surface for a general $d$-dimension. However, if we concentrate on a two-dimensional CFT with $d=2$, the geodesic length and the minimal surface have the same dimension and they are described by the same geodesic curve. Therefore, the following direct relation occurs in a three-dimensional asymptotic AdS space \cite{Park:2020nvo} 
\be         \la{conjecture:microvsmacro}
l = |x - x'| \, , \quad m R = \D_O \, ,  \quad {\rm and} \quad L(|x-x'|) = 4 G S_E  ,
\ee
where $\D_O$ indicate the conformal dimension of a local operator. On the field theory side, this relation may give us a hint about a certain connection between macroscopic and microscopic correlations.  In the late time era of the two-dimensional inflationary universe, a two-point function of a local operator evolves as
\be
\bra O(\ta, x)  \,  O(\ta,  x') \ket  &\sim&  
e^{- 2 \D_0 H \ta} \ \bra O(0, x)  \,  O(0,  x') \ket  ,
\ee
where we assume that $\e \sim H \to 0$. In this case, the two-point function at the initial time $\ta=0$ is given by
\be
\bra O(0, x)  \,  O(0,  x') \ket  = \fr{1}{|x-x'|^{2 \D_O}} ,
\ee
where $|x-x'|$ indicates the comoving distance. As a result, the microscopic correlation exponentially suppresses by $e^{- 2 \D_0 H \D\ta}$ in the exponentially inflating universe.

\section{Entanglement entropy in the braneworld model}

By using the dS boundary model in the previous section, we studied the time-dependent entanglement entropy and the microscopic two-point correlation in a two-dimensional inflationary universe. Although we exactly calculated the HRT formula beyond the RT formula, the dS boundary model has several demerits. First, it is difficult to apply the dS boundary model to higher-dimensional inflationary universes. This is because of the ambiguity of boundary conditions in calculating the higher order corrections. Second, we cannot apply the dS boundary model to other universes expanding by a power-law. In order to investigate time-dependent correlations in expanding universes by a power-law, we can apply another model called the braneworld model \cite{Park:2020jio}. In this section, we first study the time-dependent entanglement entropy in an inflationary universe of the braneworld model.

The way to obtain an inflationary universe in the braneworld model is totally different from that in the dS boundary model. To obtain an inflationary universe in the dS boundary model, we considered a time-dependent AdS space which allows a dS boundary. In the braneworld model, however, we have to prepare two static AdS spaces and assume that those two AdS spaces are bordered through a one-dimensional lower brane. In this case, the cosmological constants of two bulk spaces and the brane's tension cause a nontrivial radial motion of the brane. This radial motion is governed by the so-called junction equation. To an observer living in the brane, the brane's radial motion is reinterpreted as the time evolution of the spacetime. The big difference between two models studied here is that the bulk geometries of the braneworld model are still given by static AdS geometries. Therefore, we do not need to apply the HRT formula for calculating the time-dependent entanglement entropyn of the braneworld model. In other words, the RT formula gives us an exact result, unlike the previous dS boundary model. Another important point is that the braneworld model allows us to describe various expanding universes with matter. Therefore, it would be interesting to study the time-dependent entanglement entropy of expanding universes with various matter in the braneworld model.

\subsection{Entanglement entropy in the cutoff AdS}

Parameterizing the Poincare AdS space as
\be
ds^2 = \fr{r^2}{R^2} \ls - dt^2 + dx^2 \rs + \fr{R^2}{r^2} dr^2 ,
\ee
the boundary of the AdS space is usually located at $r=\infty$. If we put a brane as a cutoff at a finite distance and denote its radial position as $\bar{r}$, the range of the AdS space is restricted to $0 \le r \le \bar{r}$. To construct the braneworld model, we assume that there exists another AdS space beyond the cutoff. Assuming the $Z_2$ symmetry, for simplicity, under $r \to 2 \bar{r} - r$, the range of the other AdS is limited to $\bar{r} \le r \le 2 \bar{r}$. Introducing a new time coordinate $\ta$ on the brane
\be
- d\ta^2 = -  \fr{r^2}{R^2} dt^2 + \fr{R^2}{r^2} dr^2 .
\ee
where $r$ indicates the time-dependent radial position of the brane, an induced metric on the brane becomes
\be
ds_B^2 = - d\ta^2 + \fr{r^2}{R^2} dx^2 .
\ee
In this case, $\ta$ and $r/R$ correspond to the cosmological time and the scale factor of the braneworld, respectively. Denoting the tension of the brane as $\s$, the relation between $r$ and $\ta$ is determined by the following junction equation (see the details in Ref. \cite{Israel:1966rt,Park:2020jio})
\be		
\ls \fr{dr}{d\ta} \rs^2=  \ls \s^2 -  \fr{4}{R^2} \rs  \fr{r^2}{4} .
\ee
Except for the critical tension $\s_c = 2/R$, the brane usually moves in the radial direction and its velocity is determined by the junction equation. The radial motion of the brane, as mentioned before, is associated with the scale factor of the braneworld
\be
r(\ta) = R e^{H \ta}  ,
\ee
with the Hubble constant
\be			\la{result:HubbleConstant}
H = \sqrt{\fr{\s^2}{4} - \fr{1}{R^2}}   .
\ee

For simple calculation, we introduce a new coordinate $z=R^2/r$ and then represent the three-dimensional AdS metric as
\be
ds^2 &=&  \fr{R^2}{z^2} \ \ls dz^2 - dt^2 + dx^2  \rs  .
\ee
In this case, the range of the radial coordinate $z$ is restricted to $\bar{z} \le z \le \infty$ where $\bar{z}$ indicates the position of the brane. From the viewpoint of the holographic renormalization, the position of the brane corresponds to a finite UV cutoff. This setup is very similar to that of the cutoff AdS space recently studied in $T \bar{T}$-deformation \cite{Smirnov:2016lqw,Cavaglia:2016oda,McGough:2016lol,Donnelly:2018bef,Park:2018snf}. When the brane does not move, the RT formula leads to the entanglement entropy of the $T \bar{T}$-deformation. To investigate the time-dependent entanglement entropy, however, we must consider the brane moving in the radial direction. Despite this fact, the entanglement entropy calculation in the cutoff AdS space is very useful to understand the time-dependent entanglement entropy we are interested in. Therefore, we start with explaining the entanglement entropy in the cutoff AdS.

In the cutoff AdS space, if we divide the boundary space into a subsystem of $-l/2 \le x \le l/2$ and its complement, the entanglement entropy derived from the RT formula is given by
\be				\la{action:HEE2}
S_E = \fr{R}{4 G} \int_{-l/2}^{l/2} dx \ \fr{\sqrt{1 + z'^2}}{z}  ,
\ee
where the prime means a derivative with respect to $x$. This action leads to the equation of motion, which determines the configuration of the minimal surface,
\be
0 =  1+ z'^2 + z z'' .
\ee
The general solution of this equation is given by
\be
z (x) = \sqrt{c_1 - \ls c_2 + x \rs^2 } ,
\ee
where $c_1$ and $c_2$ are two integral constants.

In order to determine the exact configuration of the minimal surface, we have to fix two undetermined integral constants by imposing appropriate boundary conditions. To do so, it is worth noting that the above entanglement entropy is invariant under the parity transformation $x \to -x$. Together with this fact, the smoothness of the minimal surface requires that $z'$ must vanish at $x=0$. If we denote the value of $z(0)$ as $z_0$, $z_0$ becomes a turning point of the minimal surface at which the value of $z'$ changes its sign. The turning point gives rise to an upper bound for the range of $z$ extended by the minimal surface. The existence of such a turning point enforces $c_2 = 0$. When we calculate the area of the minimal surface, the end of the minimal surface must be identified with the entangling surface defined on the brane. This implies that we must impose another boundary condition, $\bar{z} = z \ls \pm l/2 \rs$. This additional boundary condition fixes the remaining integral constant to be
\be
c_1 = \fr{l^2}{4} + \bar{z}^2 .
\ee
As a consequence, the coordinates of the minimal surface satisfies the following circular trajectory
\be			
z^2 + x^2 = \fr{l^2}{4} + \bar{z}^2 .
\ee
Here, the ranges of $z$ and $x$ are restricted to $\bar{z} \le z \le z_0$ and $-l/2 \le x \le l/2$ respectively and the turning point appears at
\be			\la{relation:zl}
z_0 = \sqrt{\fr{l^2}{4} + \bar{z}^2} .
\ee

After plugging the obtained solution into \eq{action:HEE2}, performing the integral results in
\be			\la{result:HEE}
S_E = \fr{c_{CFT}}{6} \log \ls \fr{\sqrt{l^2 + 4 \bar{z}^2} + l}{\sqrt{l^2 + 4 \bar{z}^2} - l} \rs ,
\ee
where $c_{CFT}$ means the central charge of a two-dimensional CFT \cite{Balasubramanian1999}. When the boundary position approaches the UV regulator ($\bar{z}\to 0$), this result reproduces the well-known entanglement entropy of a two-dimensional CFT 
\be
S_E = \fr{c_{CFT}}{3} \log \fr{l}{\bar{z}} .
\ee
When we calculate the time-dependent entanglement entropy in the next section, this simple example in the cutoff AdS shows how to impose the appropriate boundary conditions in the braneworld model.

\subsection{Entanglement entropy on the moving brane}

Now, let us consider the entanglement entropy in the eternal inflationary cosmology. In the braneworld model, the  eternal inflation on the brane appears when we consider an AdS bulk geometry with a noncritical brane tension $\s \ne \s_c$. Even in this case, since the bulk geometry has nothing to do with the brane's motion, the dual geometry remains static. Although calculating the time-dependent entanglement entropy of the moving brane is almost the same as the previous calculation, there exists one big difference caused by the radial motion of the brane \cite{Park:2020jio}. When we take into account the circular trajectory of the minimal surface, the boundary condition must be modified because the boundary moves. Requiring that the end of the minimal surface must attach to the moving brane, the consistent solution is given by a function of $\ta$ and $x$ 
\be
z (\ta,x) = \sqrt{\fr{l^2}{4}+ \bar{z}(\ta)^2 - x^2} .
\ee
It is worth to noting that the time dependence of the holographic entanglement entropy in the braneworld model appears due to the time-dependent boundary condition at $\bar{z} (\ta)$. Performing the integration in \eq{action:HEE2} with the obtained time-dependent solution, the resulting entanglement entropy again yields \eq{result:HEE} with the time-dependent brane position $\bar{z}(\ta)$, instead of a constant $\bar{z}$, 
\be
\bar{z}(\ta) = R e^{- H \ta} .
\ee

Despite the fact that the same entanglement entropy again appears on the moving brane, the physical implication dramatically changes due to the time-dependence of the boundary metric. In the braneworld model, the comoving distance is not physical because of the nontrivial scale factor of the expanding universe.  The physical distance in the inflating universe is given by
\be
D(\ta)= \fr{R}{\bar{z} (\ta) } \, l  = e^{H \ta} l . 
\ee
Notice that, since the scale factor generally relies on the matter, the resulting entanglement entropy and its time dependence also crucially depend on the matter. In the inflationary universe, the resulting entanglement entropy becomes
\be			
S_E = \fr{c}{6} \log \ls \fr{\sqrt{e^{2 H \ta} l^2 + 4 R^2} + e^{H \ta} l}{\sqrt{e^{2 H \ta} l^2 + 4 R^2} -e^{H \ta} l} \rs  .
\ee
At the initial time ($\ta=0$), the entanglement entropy reduces to becomes the result of the cutoff AdS in \eq{result:HEE}.

In the late time era, the time-dependent entanglement entropy becomes perturbatively
\be
S_E = \frac{c_{CFT} }{3}  H  \ta  - \frac{ c_{CFT} }{3} \log \ls \fr{R} {l}\rs
+ \frac{c_{CFT} }{3}  \fr{R^2 }{l^2}   e^{-2 H \ta}  + {\cal O} \ls e^{-4 H \ta} \rs . 
\ee
Similar to the previous dS boundary model, the entanglement entropy of the braneworld model increases linearly with time during eternal inflation. This also implies that, similar to the previous result, the two-point function of a local operator exponentially suppresses with time
\be
\bra O(\ta,x)  \,  O(\ta,x') \ket  \sim e^{- 2 \D_O H   \ta} \bra O(0,x)  \,  O(0,x') \ket  .
\ee
where 
\be
 \bra O(0,x)  \,  O(0,x') \ket   = \fr{1}{|x - x'|^{2 \D_O}} .
\ee
This result shows that the microscopic two-point function of a local operator suppresses more rapidly as the expansion rate increases and the conformal dimension of the local operator becomes large.

\subsection{$c$-function in the braneworld model}

Let us think of the $c$-theorem of the braneworld model. In the braneworld model unlike the dS boundary model, there exist two characteristic energy scales. One is the inverse of the subsystem size which is the energy scale used in the dS boundary model. The other energy scale is represented by the radial position of the brane. In the dS boundary model, since we considered a static boundary located at the infinity of the AdS space, the boundary position does not give an any new scale. On the other hand, the boundary of the braneworld model can change its radial position which, in the usual holographic setup, is reinterpreted as the energy scale of the dual field theory. From the holographic renormalization point of view, it seems to be more natural to take into account the radial coordinate as the energy scale rather than the inverse of the subsystem size. Note that in the braneworld model, the subsystem size and the radial position of the brane are given by two independent variables depending on the initial conditions. However, the on-shell configuration of the minimal surface relates these two independent values via the time-dependent turning point $z_0 (\ta)$. Inversely, the turning point is determined by two independent variables, $\bar{z} (\ta)$ and $D(\ta)$
\be			\la{result:RelationzL}
z_0 (\ta)  = \fr{\bar{z}}{2 R}  \sqrt{ D^2 + 4 R^2} ,
\ee 
and gives rise to a unique IR energy scale. This IR energy scale is the lowest energy measured by the entanglement entropy. Therefore, it would be more reliable to identify the IR energy scale with the RG scale of the dual field theory. Then, the $c$-function in the braneworld model  is redefined in terms of $z_0$ rather than $l$ or $L$. To do so, we rewrite the entanglement entropy as a function of the time-dependent turning point, $z_0$, 
\be				\la{Result:HEEinAdS3}
S_E = \fr{c_{CFT}}{6} \log \ls  \fr{z_0   + \sqrt{z_0^2  - \bar{z}^2}}{z_0  - \sqrt{z_0^2  - \bar{z}^2}} \rs  .
\ee
For $z_0 / \bar{z} \to \infty$ which is the same limit as $D \to \infty$ in \eq{result:RelationzL}, the $c$-function reads in terms of the RG scale $z_0$
\be
c \equiv 3 \fr{\pa S_E}{\pa \log z_0}  = \fr{z_0}{\sqrt{z_0^2 - \bar{z}^2}} c_{CFT}  .
\ee
Rewriting $z_0$ in terms of the physical subsystem size $D$ in the $D \to \infty$ limit
\be
z_0 \approx \fr{\bar{z}}{2 R} \  D ,
\ee
the resulting $c$-function becomes perturbatively in the late time era
\be
c   = \lb 1 + 2   \fr{ R^2}{l^2} e^{- 2 H \ta}+ {\cal O} \ls e^{- 4 H \ta} \rs  \rb  c_{CFT}  .
\ee
This prescription is consistent with the result \eq{result:cfunctionindSmodel} obtained in the dS boundary model. Since $z_0$ in the dS boundary model monotonically increases with $l$, as shown in \eq{Result:tpvssize}, rewriting the $c$-function in terms of $z_0$ instead of $l$ does not change the qualitative behavior in \eq{result:cfunctionindSmodel}. After this prescription for the RG scale, the two models considered in this work give rise to the almost same time-dependent $c$-function which monotonically decreases with time and finally approaches the CFT result.

\section{In the expanding universe by a power-law}

Now, let us consider a braneworld model in the black hole geometry
\be
ds^2 = \fr{r^2}{R^2} \ls - f(r) dt^2 + dx^2 \rs + \fr{R^2}{r^2 f(r)} dr^2 ,
\ee
with a blackening factor given by
\be
f(r) = 1 - \fr{m}{r^2}  .
\ee
Assuming that the brane has a critical tension $\s=\s_c$, the radial motion of the brane is governed by the following junction equation 
\be		
\fr{dr}{d\ta} =  \pm \fr{\sqrt{m}}{R} .
\ee
Therefore, the radial position of the brane is given by
\be			\la{Result:braneposition}
\bar{r} =  \pm  \fr{\sqrt{m}}{R} \ta + \bar{r}_0 ,
\ee
where $\bar{r}_0$ indicates the brane's initial position. The induced metric on the brane reduces to
\be
ds_B^2 = - d \ta^2 + \fr{\bar{r}^2}{R^2} dx^2 .
\ee
Since $\bar{r}$ depends on the cosmological time,  the induced metric is an FLRW type metric. In this case, $a=\bar{r}/R$ plays a role of a scale factor and the resulting geometry represents an expanding universe with $a \sim \ta$. This expansion rate is the same as that of a two-dimensional universe with radiations.  Therefore, the braneworld in a three-dimensional black hole geometry realizes the radiation-dominated universe in the brane. Noting that the initial position $\bar{r}_0$ can be ignored for $\ta \to \infty$, the scale factor in the late time era reduces to
\be
a = \fr{\sqrt{m}}{R} \ta.
\ee

For convenience, we introduce a new coordinate $z=R^2/r$. Then, the black hole metric is rewritten as
\be
ds^2 = \fr{R^2}{z^2} \ls - f(z) dt^2 + dx^2 + \fr{1}{ f(r)} dz^2 \rs  ,
\ee
with
\be
f(r) = 1 -  \fr{z^2}{z_h^2}  .
\ee
where $z_h^2= R^4/m$. On this black hole geometry, the holographic entanglement entropy is defined as
\be
S_E = \frac{R}{4G} \int_{-l/2}^{l/2} dx \ \frac{\sqrt{z'^2 + f}}{z \sqrt{f}} .
\ee
Using the conserved quantity, the size of the subsystem is determined in terms of the turning point $z_0$
\be
l &=& - \int_0^{z_0} dz \ \frac{2 z}{\sqrt{f \left( z_0^2 -z^2 \right)}} \\ \nonumber
  &=& z_h \lb 2 \log \left( \sqrt{ z_h^2- \bar{z}^2}+\sqrt{z_0^2-\bar{z}^2} \right)   - \log \ls   z_h^2-z_0^2  \rs \rb .
\ee
The corresponding entanglement entropy becomes
\be
S_E &=& \frac{R z_0 z_h}{2 G } \int_{\e}^{z_0} dz \frac{1}{z \sqrt{z_h^2 - z^2}  \sqrt{z_0^2 - z^2}}  \\ \nonumber
&=& \frac{c}{6}  \log \left(\frac{z_0 \sqrt{z_h^2-\bar{z}^2}+z_h \sqrt{z_0^2-\bar{z}^2} }{z_0
   \sqrt{z_h^2-\bar{z}^2}- z_h\sqrt{z_0^2- \bar{z}^2} }\right)  .
\ee
In the limit of $z_h  \to \infty$, the leading part of this result reduces to the previous one \eq{Result:HEEinAdS3} 
\be
S_E = \fr{c}{6} \log \ls  \fr{z_0   + \sqrt{z_0^2  - \bar{z}^2}}{z_0  - \sqrt{z_0^2  - \bar{z}^2}} \rs
+ \fr{c z_0 }{6 z_h^2}  \sqrt{z_0^2 - \bar{z}^2} .
\ee 
Moreover, when the brane is located in the UV region, it is further reduced to
\be
S_E = \fr{c}{3} \log \fr{2 z_0}{\bar{z}} + \fr{c}{24}  \ls \fr{2 (2 z_0^2 -  \bar{z}^2)}{z_h^2} - \fr{\bar{z}^2}{z_0^2} \rs ,
\ee
where $\bar{z}$ exactly plays a role of the UV cutoff.

Now, let us consider another interesting limit, $z_0 \to z_h$. In this case, the system size is given by
\be
l = z_h  \lb \log \ls \fr{z_h^2 - \bar{z}^2}{z_h^2 - z_0^2} \rs + 2 \log 2 \rb ,
\ee
and diverges for $z_0 \to z_h$. This fact indicates that the limit of $z_0 \to z_h$ corresponds to the large subsystem size limit. Using this result, the corresponding entanglement entropy is rewritten in terms of the subsystem size $l$
\be
S_E = \fr{c}{6} \fr{l}{z_h} + \fr{c}{3}  \log \fr{z_h}{\bar{z}} + \cdots .
\ee
When the brane is located near the black hole horizon, the entanglement entropy is proportional to the volume of the subsystem rather than the area of the entangling surface. In this case, the leading term is exactly the same as the Bekenstein-Hawking entropy. In the time-independent geometry, this result shows that the entanglement entropy approaches the thermal entropy in the IR limit \cite{Kim:2016jwu,Park:2015hcz}. However, this is not case in the expanding universe. Recalling that the initial position of the brane in \eq{Result:braneposition} is negligible in the late time era, the radial position of the brane is simply approximated by 
\be
\bar{z} 
= \fr{R}{a}  ,
\ee
where $a$ denotes the scale factor $a =\ta/z_h$. In addition, remembering that  the physical distance in the expanding universe is given by $D = a l$, the entanglement entropy of the large subsystem in the late time era reduces to
\be
S_E 
&=& S_{th} + \fr{c}{3}  \log \fr{2 \ta}{R} ,
\ee
where $S_{th}= R l /(4 G z_h)$ corresponds to the thermal entropy contained in the subsystem \cite{Ryu:2006bv,Casini:2008cr,Blanco:2013joa,Park:2015hcz,Kim:2016jwu}.

It has been shown that in the time-independent black hole geometry, the quantum entanglement entropy evolves to the thermal entropy in the IR region which follows the volume law. The first term of the above result indicates such a thermal entropy corresponding to the IR entanglement entropy. On the other hand, the second term corresponds to the quantum correction which increases logarithmically with time. In the time-independent geometry, the thermal entropy contribution is usually dominated in the IR regime and leads to thermalization of  matter \cite{Kim:2016jwu}. However, the time dependence of the entanglement entropy in the expanding universe spoils thermal equilibrium in a sufficiently late time and prevents the matter from being thermalized, {\it dethermalization}. The typical time scale of dethermalization is given by
\be         \la{Result:timescale}
\ta_* \approx \fr{R}{2} e^{l/(2 z_h)} .
\ee
This implies that the dethermalization time is exponentially proportional to the subsystem size. 

In this two-dimensional radiation-dominated universe, the equal time two-point function suppresses by the following power law
\be
\bra O (\ta,x) \, O (\ta, x') \ket \sim  \ta^{- 2 \D_O} \bra O (0,x) \, O (0, x') \ket ,
\ee
where the initial two-point function is given by
\be
\bra O (0,x) \, O (0, x') \ket  &\sim& \ e^{- \D_O |x-x'|}  \quad {\rm for} \ |x - x'| \ll z_h , \nn
&\sim& \fr{1}{|x - x'|^{2 \D_O}} \quad {\rm for} \ |x - x'| \gg z_h .
\ee
As the conformal dimension of the local operator becomes larger, its two-point function decreases more rapidly.


\section{Discussion}

In the standard cosmology, it was well known that our universe expands with a different speed relying on the matter content contained in the universe. To understand the quantum features of such an expanding universe, it would be interesting to investigate the quantum entanglement entropy of an expanding universe. The entanglement entropy usually measures the macroscopic correlation between two subsystems across their border. In an expanding universe, since the background spacetime is time-dependent, one can easily expect that the entanglement entropy is also given by a time-dependent function. In this case, a nontrivial time-dependent background geometry even for a free QFT can cause a nontrivial interaction between fields. Therefore, it is not easy to evaluate the time-dependent entanglement entropy on the QFT side. In the present work, we took into account a two-dimensional QFT as a toy model and investigate how to realize time-dependent background geometries by using the holographic technique. We considered two distinct holographic models, the dS boundary and braneworld models. Although the eternal inflation was realized in both  models, it is not easy to find the dual geometry of the power-law expansion in the dS boundary model. In the braneworld model, on the other hand, the eternal inflation and the power-law expansion can be described by the motion of the brane in the AdS space without and with an appropriate bulk matter.  

It was proposed that the entanglement entropy of a strongly interacting QFT is proportional to the area of the minimal surface extending to the dual geometry. If the dual background geometry becomes time-dependent, we must exploit the HRT formula instead of the RT formula. In this work, we calculated the time-dependent entanglement entropy by applying the HRT formula and found that the RT formula gives rise to the leading contribution of the HRT formula in the late time era, which is consistent with the argument in Ref. \cite{Koh:2020rti}. The exact calculation of the time-dependent entanglement entropy is possible only for a two-dimensional dS boundary model. If we are interested in a more realistic cosmological model, we must further take into account four-dimensional universes with a power-law expansion. In general, it is not easy to generalize the dS boundary model to a higher-dimensional theory with a power-law expansion. Due to this reason, we also considered the braneworld model which can describe the power-law expansion. For the two-dimensional toy model, both models considered here give rise to the similar time-dependent entanglement entropy which increases linearly with time in the late time era.  

From the time-dependent entanglement entropy, we also investigated how the central charge varies during the time evolution of the expanding universes. In the dS boundary model, since the observation energy scale is directly connected only to the subsystem size, we studied the time-dependent $c$-function in terms of the subsystem size. As a result, the $c$-function in the inflationary universe monotonically decreases with time similar to the RG flow of the $c$-function and finally approaches the central charge of the dual CFT after infinite time. In the braneworld model unlike the dS boundary model, two dimensionful parameters specify the energy scale of the boundary theory. One is the subsystem size similar to the dS boundary model and the other is given by the position of the brane. In the holographic renormalization procedure, the energy scale of the dual QFT is characterized by the radial position of the boundary. Due to the existence of two-dimensionful parameters in the braneworld model, understanding the change of the $c$-function becomes obscure. In this work, we considered the turning point of the minimal surface, instead of the subsystem size and the brane's position, as the energy scale of the dual QFT. The reason is that the turning point determines the lowest energy scale the entanglement entropy can measure. Under this new reinterpretation, we showed that the $c$-functions of two holographic models give rise to the same qualitative features, decreasing with time and finally approaching the central charge of the dual CFT.

In the braneworld model, we also considered a brane moving in a three-dimensional AdS black hole. In the holographic setup, the AdS black hole usually gives rise to a nonvanishing boundary stress tensor which is traceless. On the dual field theory side, this traceless stress tensor represents excitation of the massless gauge boson. As a consequence, the brane's motion in the AdS black hole corresponds to the radiation-dominated universe on the brane where the equation of state parameter is given by $w=1$. In this holographic setup, we showed that a two-dimensional universe in the radiation-dominated era expands linearly with time, as it should do. We also found that the entanglement entropy in the radiation-dominated era increases by $\log \ta$. In general, a static black hole geometry allows the well-defined Hawking temperature, so that the holographic dual theory maps to a finite temperature thermal field theory. In the expanding universe, the braneworld model shows that the thermal system becomes {\it dethermalized} after the typical time scale in \eq{Result:timescale}.

For a two-dimensional dual QFT, intriguingly, the entanglement entropy is directly associated with the two-point correlation function of a local operator. This fact becomes manifest on the dual gravity side where the entanglement entropy and two-point function are described by the same geodesic length. Using this relation, we also investigated how the two-point function of a local scalar operator changes in time. In the expanding universes, we showed that the two-point function of a scalar operator suppresses exponentially for eternal inflation or by a power law in the radiation-dominated universe. In the present work, we seriously exploit the conjecture \eq{conjecture:microvsmacro} which relates the macroscopic entanglement entropy to the microscopic two-point function for a two-dimensional field theory. Although this relation looks obvious on the dual gravity side, it would be important to understand the underlying structure of this relation from the QFT viewpoint. We hope to report more results in future works. 

\vspace{1cm}

{\bf Acknowledgement}

This work was supported by the National Research Foundation of Korea(NRF) grant funded by the Korea government(MSIT) (No. NRF-2019R1A2C1006639).



\bibliographystyle{apsrev4-1}
\bibliography{References}

\end{document}